\documentclass[aps,prd,twoside,twocolumn,superscriptaddress,floatfix,nofootinbib,showpacs]{revtex4-2}

\usepackage{amsmath}
\usepackage{booktabs}
\usepackage{graphicx}
\usepackage{natbib}
\usepackage{hyperref}
\usepackage{cleveref}

\newcommand{\aap}{Astron. Astrophys.}
\newcommand{\apjl}{Astrophys. J. Lett.}
\newcommand{\jcap}{J. Cosmol. Astropart. Phys.}
\newcommand{\mnras}{Mon. Not. R. Astron. Soc.}

\begin{document}

\title{Simultaneous inference of isotropic and anisotropic polarization rotation angles for next-generation cosmic microwave background experiments}

\author{Yue Zhang}
\affiliation{Department of Astronomy,
University of Science and Technology of China, Hefei 230026, China}
\affiliation{School of Astronomy and Space Science, University of Science and Technology of China, Hefei 230026, China}

\author{Chang Feng}
\altaffiliation{Corresponding author: changfeng@ustc.edu.cn}
\affiliation{Department of Astronomy,
University of Science and Technology of China, Hefei 230026, China}
\affiliation{School of Astronomy and Space Science, University of Science and Technology of China, Hefei 230026, China}

\author{Sen Li}
\affiliation{Department of Astronomy,
University of Science and Technology of China, Hefei 230026, China}
\affiliation{School of Astronomy and Space Science, University of Science and Technology of China, Hefei 230026, China}

\author{Filipe B. Abdalla}
\affiliation{Department of Astronomy,
University of Science and Technology of China, Hefei 230026, China}
\affiliation{School of Astronomy and Space Science, University of Science and Technology of China, Hefei 230026, China}

\begin{abstract}
The ultrahigh-sensitivity polarization imaging of the cosmic microwave background (CMB) is a treasure trove for new physics. Searching for a predicted polarization angle rotation known as the cosmic birefringence effect is a pursuit of the next-generation CMB experiments, and this new polarization signal would be important for testing fundamental physics theories.
However, such a delicate rotation effect can be confused by different mechanisms including polarized Galactic foregrounds and instrumental calibration effects. Also, the cosmological rotation effects may arise from both the background evolution and spatial fluctuations of axion-like particles, leaving complicated imprints on CMB polarization fluctuations. 
In this work, we establish a comprehensive modeling of cosmological rotation effects, the instrumental miscalibration of polarization angles and complex foreground polarization with higher-order fluctuations, and perform a Bayesian analysis to infer both isotropic and anisotropic rotation angles simultaneously for foreground-mitigated polarization datasets from submillimeter to millimeter wavelengths. 
We find that such a Bayesian formalism can effectively extract the rotation effects for the next-generation CMB experiments and investigate the impact of the known sources including Galactic foregrounds and lensing on the inferred cosmological signals.
The method in this work is well suited for future polarization data analyses, and the inferred rotation effects may shed light on the nature of the axion-like particles.
\end{abstract}
\maketitle

\section{Introduction}
\label{sec:intro}

The polarization patterns of the cosmic microwave background (CMB) were formed due to Thomson scattering with the free electrons in the early Universe. Propagating in a cosmological distance, the CMB polarization might have been slightly rotated in a space with axion-like particles (ALPs)~\cite{sean}. Such a rotation is known as the cosmic birefringence (CB) effect, which is being searched for by both space-borne and ground-based experiments~\cite{bkrot0,chinarot,pbani,bkcb,actcb2,fgpolspatial}. If detected, it would have profound significance to fundamental physics. 

The CB effect can be theoretically realized by introducing a Chern-Simons coupling between the electromagnetic field and a pseudoscalar field~\cite{sean}. Such a theory indicates that the CMB polarization is dynamically rotated by ALPs with different masses~\cite{alpmasses}. The ALP-induced rotations can convert the electric-like modes ($E$-mode) to the curl-like modes ($B$-mode), thereby creating a nonzero $EB$ cross-power spectrum. Similarly, the Faraday rotation of primordial magnetic fields can also create a nonzero $EB$ power spectrum but it is frequency-dependent and there has been no evidence for such a signal so far~\cite{pmf}. If the ALPs are assumed to be dark-matter-like, the rotation effects would be tomographic~\cite{alpmasses} and may even be dynamical in the time domain~\cite{alptime}. However, if the ALPs are assumed to be dark-energy-like, the ALPs mainly change the amplitude of the $EB$ cross-power spectrum but with a slight phase shift as found in recent work~\cite{alpmasses}. Thus, the rotation effect can be approximately described by an isotropic angle which is just determined by the ALP field values at the recombination and today.

In addition to those cosmological origins, the polarization rotation effects can also be generated by miscalibration of detectors' polarization angles (hereafter referred to as miscalibration angle). The nonzero $EB$ power spectra generated by this mechanism form the basis to internally calibrate the polarization angles~\cite{selfcal}. The CMB miscalibration angles are strongly degenerate with the cosmological CB effects, although such a degeneracy can be physically broken by massive ALPs~\cite{alpmasses}. It is a great challenge to precisely decouple these two angles from the current or future observational datasets. 
A recent work used the foreground polarization to isolate the polarization signals due to the instrumental effects and found evidence of a 0.3 degree rotation angle from the Planck polarization data~\cite{mk} (hereafter, we refer to this method as the MK method).  
Moreover, a follow-up work has achieved a much higher detection significance with a joint dataset combining multifrequency CMB observations from the Planck satellite and the Wilkinson Microwave Anisotropy Probe (WMAP)~\cite{followup1, followup2}.
In addition to the measurements from space, the Atacama Cosmology Telescope (ACT) recently applied the MK method to its data release six and detected a nonzero isotropic rotation angle at 0.2 degrees~\cite{actcb}. Both data analyses with the MK method showcased that inclusion of the foreground polarization can effectively break the degeneracy between the polarization angle miscalibration and the cosmological angles for current CMB datasets. 

The next-generation CMB experiments, such as the Simons Observatory (SO) and the South Pole Telescope (SPT)~\cite{so, spt3g}, will map the polarized sky with extremely high sensitivity and resolution from submillimeter to millimeter wavelengths. Sophisticated foreground removal procedures become indispensable and must be applied to the multifrequency datasets to meet data quality requirements for an array of scientific objectives. Thus, it is necessary to develop an extended approach to address these issues. 
The MK method can decouple the miscalibration angles from the cosmological ones by taking advantage of the scaling-law feature of the polarized foreground power spectra~\cite{mk}. Mitigating the foreground polarization, on which the decoupling of the cosmological and instrumental angles depends, breaches the basis of the MK method, leading to a strong degeneracy among the isotropic angles again. This is the case for most of the ground-based CMB experiments which have optimized the scan strategies to avoid sky patches with highly contaminated foregrounds. Conceivably, the capability of extracting the cosmological angles would be compromised with foreground polarization mitigated. However, given the fact that the next-generation CMB experiments will have sufficiently high sensitivities, the loss of constraining power due to the reduced polarization foreground may be compensated by improving the CMB polarization sensitivities with better resolved acoustic peaks, so the MK approach may still be effective.

Meanwhile, the polarized foreground is highly non-Gaussian as revealed by recent Planck observations~\cite{fgpolspatial}. The foreground non-Gaussianity lies in two aspects, i.e., the intrinsic amplitude fluctuations and direction-dependent spectral energy distributions (SEDs) with the latter investigated in recent works~\cite{cmilc,ocmilc,2025arXiv251207235L}. Higher-order polarization fluctuations can be generated by these non-Gaussianities and would further complicate the MK approach, in which the foreground polarization plays a crucial role. This could be a significant issue for the high-sensitivity CMB polarization observations in the future, especially for the ground-based CMB experiments which normally scan small sky patches where the polarized foregrounds possess more spatial variations.

In addition to the isotropic rotation effects induced by the Chern-Simons coupling between the electromagnetic and pseudoscalar fields, the associated spatial fluctuations of the scalar field can also create anisotropic cosmic birefringence (ACB) effects, leaving unique scale-dependent perturbations on the CMB power spectra~\cite{mingzhe08}. Moreover, if there were certain correlations between the isotropic and anisotropic rotation angles, the constraints on both angles would be more complicated. 

In this work, we introduce a comprehensive model consisting of three types of rotation effects aforementioned and foreground polarization with higher-order effects. We adopt a forward modeling of these signals at the power-spectrum level and investigate a simultaneous inference of different rotation angles and foreground parameters from future high-sensitivity and multifrequency polarization measurements as motivated by recent works~\cite{keating, christian}.

The main differences between the MK formalism and this work lie in a few aspects. First, we incorporate the anisotropic rotation effects in addition to the isotropic rotation angles from cosmological and instrumental origins. If correlated, the anisotropic rotation angles could contribute partially to the isotropic ones, potentially complicating the detection. Second, we perform a foreground removal procedure and obtain foreground-mitigated polarization data, with which we decouple the cosmological rotation angle $\beta$ from the angles $\alpha$ arising from the miscalibration angles. In contrast, the MK formalism requires substantial foreground polarization signals and takes the differenced power spectra $C_{\ell}^{EE}-C_{\ell}^{BB}$ to cancel out any modeling mismatch of the foreground power spectra to avoid foreground leakages. Third, we also consider the higher-order fluctuations in the foreground polarization. 

This paper is organized as follows. In Sec.~\ref{sec:model}, we introduce the CMB and foreground models investigated in this work. In Sec.~\ref{sec:method}, we describe the likelihood function and the parameter estimation method. In Sec.~\ref{sec:result}, we present our results. We conclude in Sec.~\ref{sec:conclusion}.

\section{Power spectrum models with different rotation effects}
\label{sec:model}

In this work, we extend the MK method and investigate the inference of different rotation angles by including the anisotropic rotation angles and the higher-order foreground polarization. 

\subsection{CMB polarization power spectra with different rotation effects}
\label{subsec:cmb}

We extend the non-perturbative calculations described in~\cite{Mingzhe_Li_2013} to include the instrumental isotropic rotation effects. The CMB polarization at the direction $\hat{n}$ after being rotated by an angle $\alpha_i(\hat{n})$ is $[Q_i\pm \mathrm{i}U_i](\hat{n})$, which can be described as 
\begin{equation}
    [Q_i\pm \mathrm{i}U_i](\hat{n}) \equiv [\tilde{Q} \pm \mathrm{i}\tilde{U}](\hat{n})e^{\pm2\mathrm{i}\alpha_i(\hat{n})},
    \label{eqn:rotated-stokes}
\end{equation}
where the tilde symbols denote the primordial CMB polarization, and the index $i$ refers to a frequency band. The rotation angle $\alpha_i(\hat{n})$ can be decomposed into three parts:
\begin{equation}
    \alpha_i(\hat{n}) \equiv \beta + \alpha_i + \delta\alpha(\hat{n}).
    \label{eqn:angle-decompose}
\end{equation}
Here, $\beta$ refers to the isotropic cosmic birefringence angle and is frequency independent. $\delta\alpha(\hat{n})$ refers to the anisotropic cosmic birefringence angle at the direction of $\hat{n}$, with zero mean and root-mean-square (RMS) quantified by the two-point correlation function, i.e., the angular power spectrum. The symbol $\alpha_i$ refers to the miscalibration angle at frequency $\nu_i$.

The CMB polarization can be expanded with spherical harmonics as $ [Q\pm {\rm i}U](\hat{n}) = \sum_{\ell m}{}_{\pm2}a_{\ell m}\ {}_{\pm2}Y_{\ell m}(\hat{n})$, where the coefficients can be further decomposed into $E$- and $B$-modes via ${}_{\pm2}a_{\ell m}=E_{\ell m}\pm \mathrm{i}B_{\ell m}$. We can use the real-space correlation functions to derive the power spectra affected by three types of rotation effects~\cite{corr1, corr2, Mingzhe_Li_2013} and the details can be found in Appendix~\ref{appsec:derivation-cb}. The rotated power spectra can be expressed as  
\begin{eqnarray}
    {C}_{\ell}^{{\rm rot}, EE, (ij)} &-& {C}_{\ell}^{{\rm rot}, BB, (ij)} \nonumber\\
	  &=& \cos(4\beta + 2\alpha_i + 2\alpha_j) e^{-4C^{\alpha}(0)}\nonumber\\
	     &&\sum_{\ell^{\prime}}\frac{2\ell^{\prime} + 1}{2}
	     ({\tilde C}_{\ell^{\prime}}^{EE} - {\tilde C}_{\ell^{\prime}}^{BB})\nonumber\\
	     &&\int_{-1}^{1} d_{-22}^{\ell^{\prime}}(\theta) d_{-22}^{\ell}(\theta)
	     e^{-4C^{\alpha}(\theta)}\mathrm{d}\cos\theta ,
	     \nonumber\\
    {C}_{\ell}^{{\rm rot}, EB, (ij)} &+& C_\ell^{{\rm rot}, EB, (ji)}\nonumber\\
	  &=& \sin(4\beta + 2\alpha_i + 2\alpha_j) e^{-4C^{\alpha}(0)}\nonumber\\
      &&\sum_{\ell^{\prime}}\frac{2\ell^{\prime} + 1}{2}
	     ({\tilde C}_{\ell^{\prime}}^{EE} - {\tilde C}_{\ell^{\prime}}^{BB})\nonumber\\
	     &&\int_{-1}^{1} d_{-22}^{\ell^{\prime}}(\theta) d_{-22}^{\ell}(\theta)
	     e^{-4C^{\alpha}(\theta)}\mathrm{d}\cos\theta.
	     \label{polrotated}
\end{eqnarray}
Here, the Wigner d-function is $d_{mm'}^{\ell}(\theta)$ and the real-space correlation function of the anisotropic rotation field is 
\begin{equation}
    C^{\alpha}(\theta) \equiv \left<\delta\alpha(\hat{n}) \delta\alpha(\hat{n^\prime})\right> = \sum_L \frac{2L + 1}{4\pi} C_L^{\alpha\alpha} P_L(\cos\theta), 
\end{equation}
where we define $\theta$ as the angle between the direction $\hat{n}$ and $\hat{n^\prime}$, thus we have $\cos\theta = \hat{n}\cdot\hat{n^\prime}$. Also, the power spectrum of the anisotropic rotation field is $C_{L}^{\alpha\alpha}\delta_{LL'}\delta_{MM'}=\langle\alpha^{\ast}_{LM}\alpha_{L'M'} \rangle$ and $\alpha_{LM}$s are spherical harmonics from $\delta\alpha(\hat n)=\sum_{LM}\alpha_{LM}Y_{LM}(\hat n)$. The Legendre function $P_{L}(\theta)$ is related to the d-function as $P_{L}(\cos\theta)=d_{00}^{L}(\theta)$.

In addition to the rotation effects, the CMB is also lensed by the intervening large-scale structure (LSS). Thus, the unlensed power spectra ${\tilde C}_{\ell}^{EE}$ and ${\tilde C}_{\ell}^{BB}$ in Eq.~(\ref{polrotated}) are replaced by the lensed ones. We use the public code \texttt{CAMB}~\cite{Lewis:1999bs} to compute the CMB power spectra with the Planck 2018 best-fit cosmological parameters~\cite{2020A&A...641A...6P}. We also introduce an amplitude parameter $A_{\rm lens}$ to consider the impact of lensing effects. Specifically, we adjust the lensed power spectra according to the relations as $C_{\ell}^{EE} ={\tilde C}_{\ell}^{EE, 0}+A_{\rm lens} (C_{\ell}^{EE, 0}-{\tilde C}_{\ell}^{EE, 0})$ and $C_{\ell}^{BB} = A_{\rm lens} C_{\ell}^{BB, 0}$ without considering any primordial $B$ modes. Here, the superscript 0 denotes the input power spectra directly calculated by \texttt{CAMB}.

The theoretically predicted rotation power spectrum is scale-invariant if a massless scalar field is assumed~\cite{alpmassless}. However, we treat the rotation power spectrum as agnostic and assume a piece-wise power spectrum instead of the scale-invariant one as normally adopted in the literature. Specifically, we consider two scenarios. The first scenario is referred to as \textbf{LOWL} where the first few moments of the rotation fields, i.e., $L(L+1)/(2\pi)C_L^{\alpha\alpha}=A^{\alpha}_{k}$ for multipoles $L=1, 2, 3$ are inferred from the mock data. The second scenario is referred to as \textbf{HIGHL} and we assume a piece-wise band power with smaller angular scales as follows
\begin{equation}
    \frac{L(L+1)}{2\pi}C_L^{\alpha\alpha} = 
    \begin{cases}
        A^{\alpha}_{1}, & 1 \le L \le 4 \\
        A^{\alpha}_{2}, & 5 \le L \le 10 \\
        A^{\alpha}_{3}, & 11 \le L \le 50 \\
    \end{cases},
    \label{eqn:claa}
\end{equation}
where \(A^{\alpha}_{k}\) \((k\in\{1, 2, 3\})\) are three free parameters. The number of piece-wise bands can be extended, but for simplicity, we only consider three bands.

From Eq.~(\ref{polrotated}), we find that the power spectra can be reduced to the standard scenario~\cite{2020PTEP.2020j3E02M} when the anisotropic birefringence effects vanish, i.e., $C_L^{\alpha\alpha} \to 0$. 

\subsection{Polarized foreground power spectra with complex spatial variations and miscalibration angles}
\label{subsec:foregrounds}

Planck has measured the $E$- and $B$-mode power spectra of the polarized foregrounds at high frequencies~\cite{pkfgpol}. The cross-power spectra $TB$ and $EB$ from these high frequency bands are found to be nonzero, indicating a complex nature of the polarized foregrounds. Also, it is found that the ratio $C_{\ell}^{fg, EE}/C_{\ell}^{fg, BB}$ is approximately a factor of two. 

Therefore, we set the foreground $EE$ power spectra to be $C_{\ell}^{fg, EE} = 2C_{\ell}^{fg, BB}$ and set the intrinsic $EB$ cross-power spectra of the polarized foregrounds to be 
\begin{equation}
    C_{\ell}^{fg, EB} = \mathcal{Z}C_{\ell}^{fg, EE} = \eta_{\ell}\sin4\psi_{\ell}C_{\ell}^{fg, EE}
    \label{eb}
\end{equation} 
following the model in~\cite{pkfgpol}. Here, $\eta_{\ell}$ is the foreground amplitude and $\psi_{\ell}$ is determined by the cross-power spectra of $TE$ and $TB$. We should note that recent studies consider $\mathcal{Z}$ as a scale-dependent parameter~\cite{followup1} but we choose $\mathcal{Z}=0.5$ as a conservative approach. At leading order, the foreground $E$- or $B$-mode power spectra can be modeled as 
\begin{equation}
    \frac{\ell(\ell+1)}{2\pi}C_\ell^{cc} 
      = A_c \left(\frac{\ell}{\ell_0}\right)^{\alpha_c} 
\end{equation}
for both the dust and the synchrotron. Here, $c$ refers to either dust ($D$) or synchrotron ($S$), $A_c$ is the foreground amplitude, $\ell_0$ is a pivot multipole and $\alpha_c$ refers to the foreground spectral index. In this work, the pivot angular scale is $\ell_0 = 80$. The power spectra of the polarized foregrounds at different frequencies thus can be modeled by incorporating the SEDs 
\begin{align}     
    \bar{S}_{\nu}^D 
      &= 
         \left(\frac{\nu}{\nu^D_{0}}\right)^{\bar{\beta}^D + 1}
         \frac{1}{\exp(h\nu/k_BT_D) - 1} 
      \label{eqn:s-dust-nu},\\
    \bar{S}_{\nu}^S 
      &=  \left(\frac{\nu}{\nu^S_{0}}\right)^{\bar{\beta}^S},
      \label{eqn:s-sync-nu}
\end{align}
where $\bar{\beta}^c$s are the spectral indices, $\nu_0^c$s are the reference frequencies, $T_D$ is the dust temperature, $h$ is the Planck constant, and $k_B$ is the Boltzmann constant. We also convert the brightness temperature to the CMB thermodynamic temperature for the SEDs. 

Planck measurements have revealed that the foreground parameters of the SEDs may not be constant and can be spatially varying~\cite{fgpolspatial}. The recent analysis investigated the impact of these spatial variations on the $B$-mode inference and adopted the constrained moment internal linear combination method to mitigate the non-negligible effects of these spatial variations~\cite{cmilc, 2025arXiv251207235L}. By assuming a simple power law, the spatial components of the spectrum indices can be modeled as $C_\ell^{\beta_c} = B_c (\ell/\ell_0)^{\gamma_c}$. In this work, we only consider the amplitude parameter $B_c$ as a free parameter and fix $\{\gamma_D, \gamma_S\} = \{-3.5, -2.5\}$ to avoid strong degeneracies. 

Following~\cite{2025arXiv251207235L} and~\cite{2021JCAP...05..047A}, the foreground $B$-mode power spectra with higher-order perturbations at frequencies $\nu_i$ and $\nu_j$ are given by
\begin{equation}
    C_{\ell}^{fg, BB, (ij)} = C_{\ell}^{ij}|_{0\times0} + C_{\ell}^{ij}|_{1\times1} + C_{\ell}^{ij}|_{0\times2},
\end{equation}
where
\begin{equation}
    \begin{aligned}
        C_\ell^{ij}|_{0\times0} 
            &= \sum_{c}\bar{S}_{\nu_i}^{\mathrm{c}}\bar{S}_{\nu_j}^{\mathrm{c}}C_{\ell}^{\mathrm{cc}}, \\            
        C_\ell^{ij}|_{1\times1} 
            &= \sum_{c} \partial_{\beta_c}\bar{S}_{\nu_i}^c\partial_{\beta_c}\bar{S}_{\nu_j}^c\sum_{\ell_1\ell_2}\frac{(2\ell_1+1)(2\ell_2+1)}{4\pi} \\
            &\quad\ \times\begin{pmatrix}\ell&\ell_1&\ell_2\\0&0&0\end{pmatrix}^2C_{\ell_1}^{cc}C_{\ell_2}^{\beta_c},\\
        C_\ell^{ij}|_{0\times2} 
            &= \sum_{c} \frac{1}{2}\left(\bar{S}_{\nu_i}^{c}\partial_{\beta_c}^{2}\bar{S}_{\nu_j}^{c}+\bar{S}_{\nu_j}^{c}\partial_{\beta_c}^{2}\bar{S}_{\nu_i}^{c}\right)C_{\ell}^{cc}\sigma_{\beta_{c}}^{2}.
    \end{aligned}
    \label{polfg}
\end{equation}
Here, $\sigma^2_{\beta_c}=\sum_\ell (2\ell + 1)/(4\pi) C_\ell^{\beta_c}$ which can quantify the level of spatial variations. The definitions and fiducial values of the SED parameters are listed in Table~\ref{table:fg-params}, and the fiducial values of the other foreground parameters can be found in Table~\ref{table:priors}.

\begin{table}[htbp]
\centering
\caption{Definition and fiducial values of the SED parameters. The fiducial values are taken from~\cite{2025arXiv251207235L}.}
\label{table:fg-params}
\begin{tabular}{ccc}
\toprule
Symbol            & Fiducial value & definition \\
\midrule
$T_{{D}}$         & $20\rm K$      & dust temperature\\
$\nu^D_{0}$       & $353\rm GHz$   & dust reference frequency\\
$\nu^S_{0}$       & $23\rm GHz$    & synchrotron reference frequency\\
$\bar{\beta}^{D}$ & $1.54$         & dust spectral index\\
$\bar{\beta}^{S}$ & $-3$           & synchrotron spectral index\\
\bottomrule
\end{tabular}
\end{table}

The CMB detectors receive the polarization signals from both the CMB and the foregrounds. In principle, both the polarization signals can be rotated by the CB angles and the miscalibration angles. However, the polarized foregrounds are mainly rotated by the detectors' miscalibration angles since the CB rotation within the Galactic scale is negligible. Thus, the rotated power spectra of the polarized foregrounds are mainly driven by the miscalibration angles via~\cite{2022A&A...662A..10E} 
 \begin{eqnarray*}
     {\hat C}_{\ell}^{fg, EE, (ij)}
       &=& \epsilon C_{\ell}^{fg, EE, (ij)}[\cos2\alpha_i\cos2\alpha_j\nonumber\\
       &&- \mathcal{Z}^{(\rm auto)}\sin(2\alpha_i+2\alpha_j)] \nonumber\\
       &&+ \epsilon C_{\ell}^{fg, BB, (ij)} \sin2\alpha_i\sin2\alpha_j, \nonumber\\
     {\hat C}_{\ell}^{fg, BB, (ij)}
       &=& \epsilon C_{\ell}^{fg, EE, (ij)}[\sin2\alpha_i\sin2\alpha_j\nonumber\\
       &&+ \mathcal{Z}^{(\rm auto)}\sin(2\alpha_i+2\alpha_j)] \nonumber\\
       &&+ \epsilon C_{\ell}^{fg, BB, (ij)}\cos2\alpha_i\cos2\alpha_j,\nonumber\\
     {\hat C}_{\ell}^{fg, EB, (ij)}
       &=& \epsilon C_{\ell}^{fg, EE, (ij)}\cos2\alpha_i\sin2\alpha_j \nonumber\\
       &&- \epsilon C_{\ell}^{fg, BB, (ij)}\sin2\alpha_i\cos2\alpha_j \nonumber\\
       &&+ \epsilon C_{\ell}^{fg, EB, (ij)}\cos(2\alpha_i+2\alpha_j).
      \label{polfgrot}
 \end{eqnarray*}

Here we introduce a scaling factor $\epsilon=0.1$ to quantify the overall foreground-removal efficiency, with the baseline value chosen following the recent BICEP/Keck measurements~\cite{bk}. We also neglect the negligible intrinsic $EB$ correlations entering the foreground $EE$ and $BB$ auto-power spectra by setting $\mathcal{Z}^{(\rm auto)}=0$, while retaining the intrinsic $EB$ contribution in the $EB$ power-spectrum model. We investigate three representative foreground-removal efficiencies, namely $0.01\epsilon$, $0.1\epsilon$, and $1\epsilon$. Although the residual foreground level is parameterized by a single overall scaling factor $\epsilon$, the underlying foreground model remains considerably more flexible, incorporating separate dust and synchrotron components with independent amplitudes ($A_D$, $A_S$), angular power-spectrum indices ($\alpha_D$, $\alpha_S$), higher-order spatial-variation parameters ($B_D$, $B_S$), frequency-dependent spectral energy distributions, and intrinsic foreground $EB$ polarization. The parameter $\epsilon$ therefore controls only the overall residual foreground level after component separation, while the remaining foreground parameters span the dominant physical uncertainties and degeneracies within the adopted power-spectrum framework.

\begin{figure*}[htp]
    \centering
    \includegraphics[width=1\linewidth]{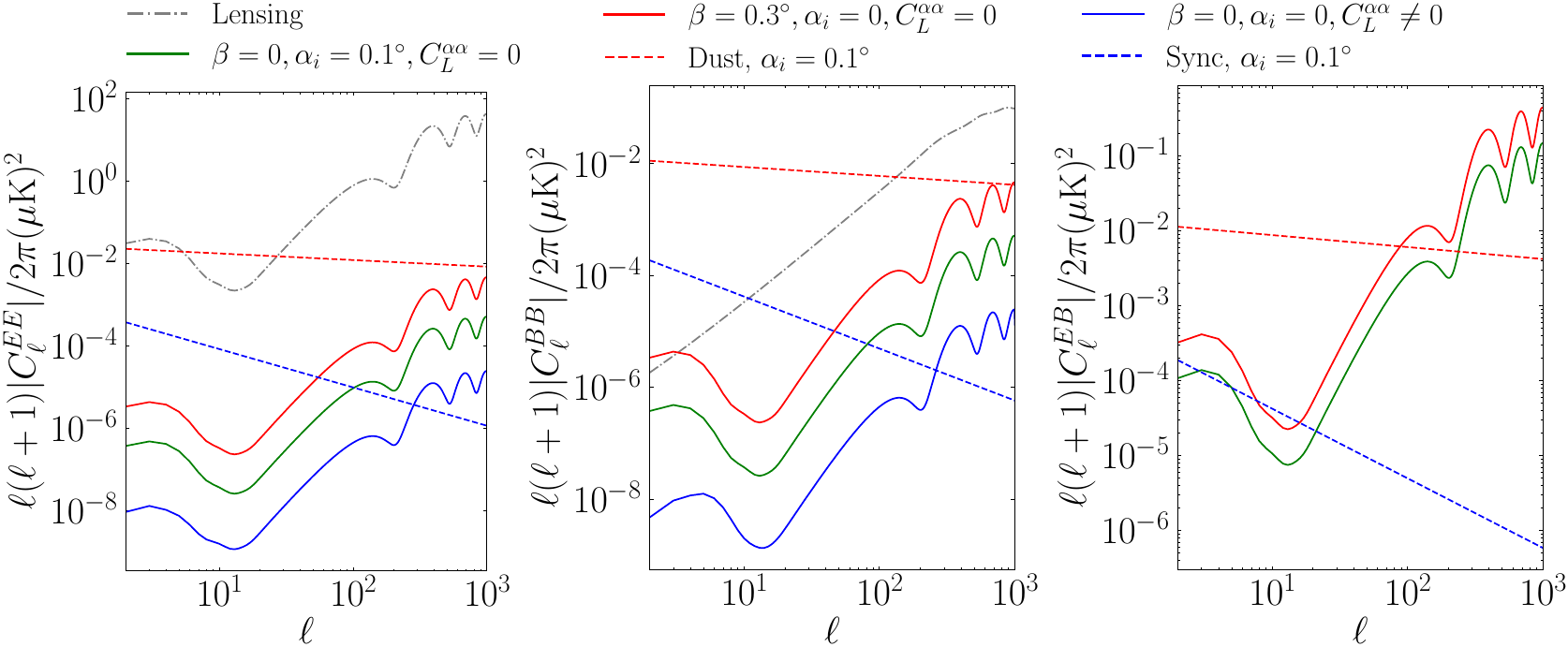}
    \caption{Power-spectrum components due to the rotation effects including cosmic birefringence and instrumental miscalibration of the polarization angles. We show the rotated power spectra for the CMB and polarized foregrounds in solid and dashed lines, respectively. The benchmark foreground level is taken from the BICEP/Keck measurements~\cite{bk} and is specifically referred to as ``$1\epsilon$" in the text. We have subtracted the unrotated but lensed components (the gray dot-dashed lines) from the $E$ and $B$ mode power spectra and show the rotation-induced contributions with absolute values for the $EE$ and $BB$ power spectra.}
    \label{fig:power-all}
\end{figure*}

\subsection{Simulations of different rotation effects at power-spectrum level}
\label{subsec:simulated-data}

In this work, we only consider five frequency bands $\{20, 50, 95, 150, 195\}$ GHz but our analysis can be extended to other frequency ranges. We choose the fiducial values listed in Table~\ref{table:priors} and generate the rotated CMB and foreground power spectra following the discussions in Sec.~\ref{subsec:cmb} and Sec.~\ref{subsec:foregrounds}. For each frequency band, we assign a miscalibration angle $\alpha_i$ and assume that the cosmological isotropic rotation angle $\beta$ is the same for all bands. 

We assume a frequency-dependent white noise component for each frequency band and a Gaussian beam profile with a FWHM $\sigma$~\cite{2002ApJ...574..566H}. The beam-deconvoluted noise power spectra for the polarization are 
\begin{equation}
    N_\ell^{EE, (ij)} = N_\ell^{BB, (ij)} =\delta_{ij}\Delta_p^2 e^{\ell(\ell+1) \sigma_i^2/(8\ln 2)}.
\end{equation}
Here, we assume the noise level is $\Delta_p = 1\rm \mu {\rm K}\mbox{-} {\rm arcmin}$ and $\sigma_i=((600\mathrm{GHz})/\nu_i)\ {\rm arcmin}$. 

The mock power spectrum measured at two frequencies $\nu_i$ and $\nu_j$ can be expressed as
\begin{eqnarray}
    {\hat C}_{\ell}^{XY, (ij)} 
      &=& C_{\ell}^{rot, XY, (ij)}(\beta, \alpha_i, \alpha_j, \{A^{\alpha}_k\}) \nonumber \\
      &&+ {\hat C}_{\ell}^{fg, XY, (ij)}(\alpha_i, \alpha_j) + N_\ell^{XY, (ij)} ,
    \label{mockdata}      
\end{eqnarray}
where $X, Y\in \{E, B\}$. There are $15$ different $EE$ and $BB$ power spectra for five frequency bands, and 25 unique $EB$ power spectra. Thus, we can construct a data vector with 55 band powers for each multipole and the data vector is defined as $\hat{\boldsymbol{x}}_{\ell} = (\hat{C}_{\ell}^{EE}, \hat{C}_{\ell}^{BB}, \hat{C}_{\ell}^{EB})$. 

In Figure~\ref{fig:power-all}, we show a representative plot of different rotation effects at $150$ GHz. In this plot, we show the power-spectrum component for the isotropic rotation angle $\beta$ in a red solid line, the miscalibration angle $\alpha_i$ in green and the anisotropic rotation $\delta\alpha$ in blue. Three different rotation angles manifest a similar power-spectrum shape, indicating there are certain degeneracies among these signals. The gray dot-dashed line denotes the lensed power spectra for comparison.
The colored dashed lines denote the power spectra of the polarized foreground rotated by the miscalibration angles since both the unrotated and rotated ones have the same power-spectrum shape, which is obviously different from that of the CMB signals. This feature makes the miscalibration angle $\alpha_i$ at each frequency separable from the observed CMB power spectra. From the $EE$ and $BB$ power spectra, the isotropic angle $\beta$ and anisotropic rotation angles $\delta\alpha$ are still degenerate. However, the $EB$ cross-power spectrum has a very negligible component from the anisotropic rotation and is mainly affected by the isotropic rotation angles so the inclusion of the $EB$ cross-power spectrum can help break the degeneracy between $\beta$ and $\delta\alpha$, making the three types of rotation angles separable from high precision polarization data. 

As seen from the rightmost panel, we assume non-negligible intrinsic $EB$ cross-power spectra that are more significant than the corresponding Planck measurements of the polarized foregrounds since we adopt a conservative approach as discussed in Sec. \ref{subsec:foregrounds}. 

We also assume that the foreground removal procedures have been applied to the polarization data and the dashed lines represent an example of the foreground residuals which are simply quantified by a parameter $\epsilon$ since we only build the mock data at the power-spectrum level in this work. We choose the benchmark foreground-removal efficiency according to the recent BICEP/Keck measurements~\cite{bk}, but adjust the efficiency parameter $\epsilon$ to test the inference of different components at different foreground-removal levels. 

\begin{figure*}[ht]
    \centering
    \includegraphics[width=1\linewidth]{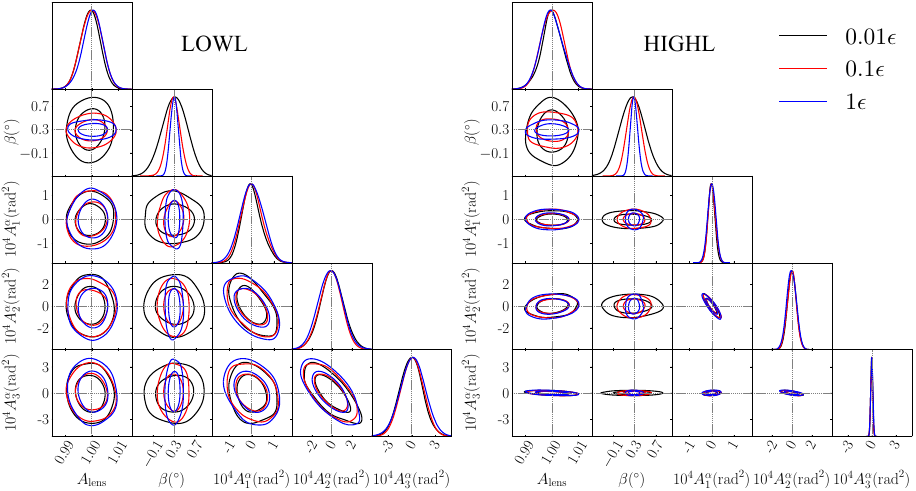}
    \caption{Posterior distributions of the cosmological parameters $\{A_{\rm{lens}}, \beta, A_{1}^{\alpha}, A_{2}^{\alpha}, A_{3}^{\alpha}\}$ for the \textbf{LOWL} and \textbf{HIGHL} cases in the left and right plots, respectively. We test three different foreground levels with $0.01\epsilon$ in black, $0.1\epsilon$ in red, and $1\epsilon$ in blue.}
    \label{fig:posterior1}
\end{figure*}

\begin{figure}[ht]
    \centering
    \includegraphics[width=1\linewidth]{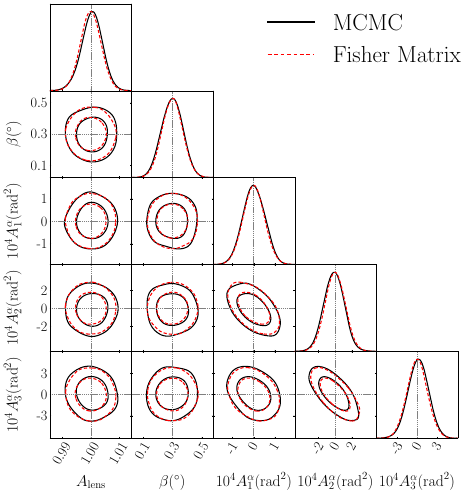}
    \caption{We cross-check the posterior distributions of the cosmological parameters $\{A_{\rm{lens}}, \beta, A_{1}^{\alpha}, A_{2}^{\alpha}, A_{3}^{\alpha}\}$ with Fisher Matrix calculations for the \textbf{LOWL} case at a 1$\epsilon$ foreground level. Both MCMC and Fisher Matrix formalism can generate consistent results.}
    \label{fig:fisher}
\end{figure}

\begin{figure}[ht]
    \centering
    \includegraphics[width=1\linewidth]{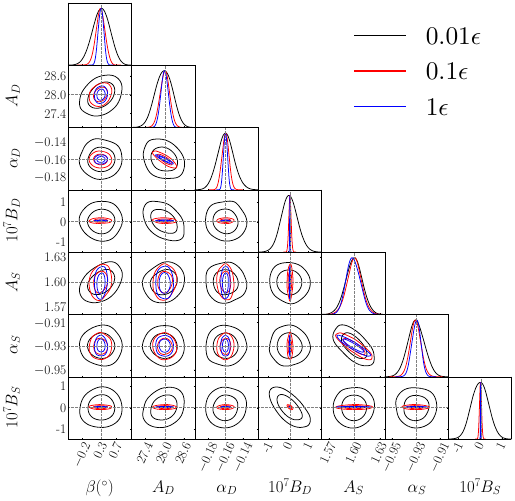}
    \caption{Posterior distributions of foreground parameters and cosmological isotropic rotation angle \{$\beta$, $A_D$, $\alpha_D$, $B_D$, $A_S$, $\alpha_S$, $B_S$\} for the \textbf{LOWL} case. We also test three foreground contamination levels with $0.01\epsilon$ in black, $0.1\epsilon$ in red, and $1\epsilon$ in blue.}
    \label{fig:posterior5}
\end{figure}

\begin{figure}[ht]
    \centering
    \includegraphics[width=1\linewidth]{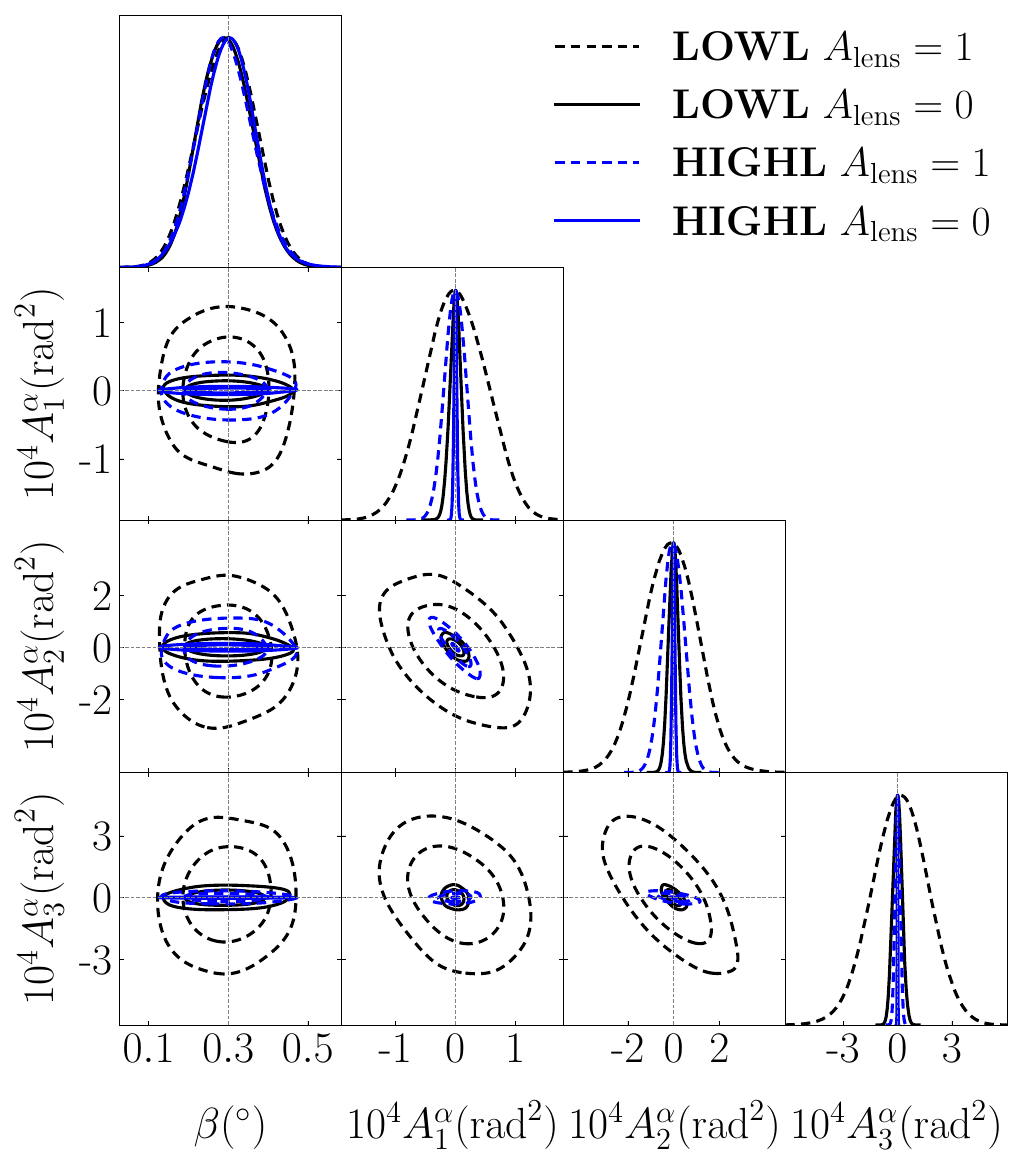}
    \caption{The impact of lensing signals on the cosmological parameters of the rotation angles. We compare the posterior distributions for $A_{\mathrm{lens}} = 1$ and $A_{\mathrm{lens}} = 0$ at $1\epsilon$ foreground level. The prior of $A_{\mathrm{lens}}$ is set to uniform distribution in the range $[-1, 1]$ when calculating the $A_{\mathrm{lens}}=0$ cases.}
    \label{fig:posterior-beta_A_CB1_A_CB2_A_CB3_BOTH_LOWL_HIGHL_Alens1_Alens0}
\end{figure}

\begin{figure*}[ht]
    \centering
    \includegraphics[width=0.95\linewidth]{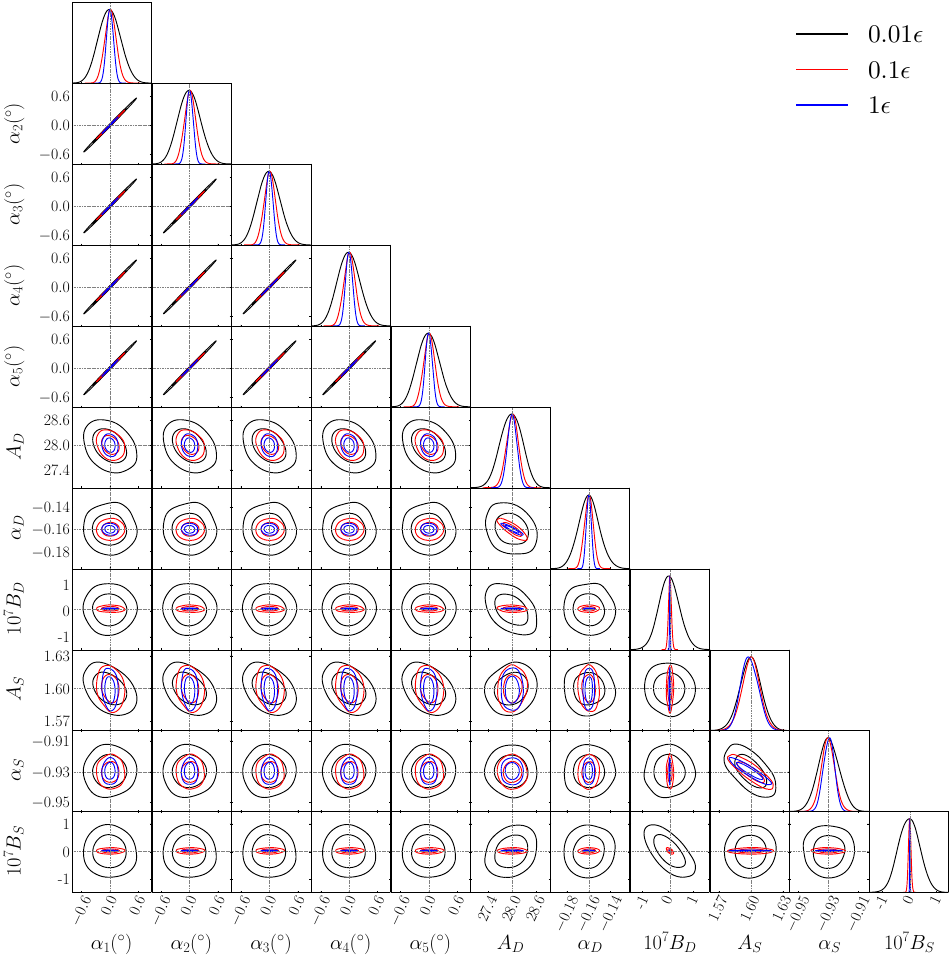}
    \caption{Posterior distributions for the miscalibration angles and the foreground parameters \{$\alpha_1$, $\alpha_2$, $\alpha_3$, $\alpha_4$, $\alpha_5$, $A_D$, $\alpha_D$, $B_D$, $A_S$, $\alpha_S$, $B_S$\}. In this test, we adopt a mock dataset with a combination of the \textbf{LOWL} and $A_{\mathrm{lens}}=1$.}
    \label{fig:posterior6}
\end{figure*}

\begin{figure*}[ht]
    \centering
    \includegraphics[width=1\linewidth]{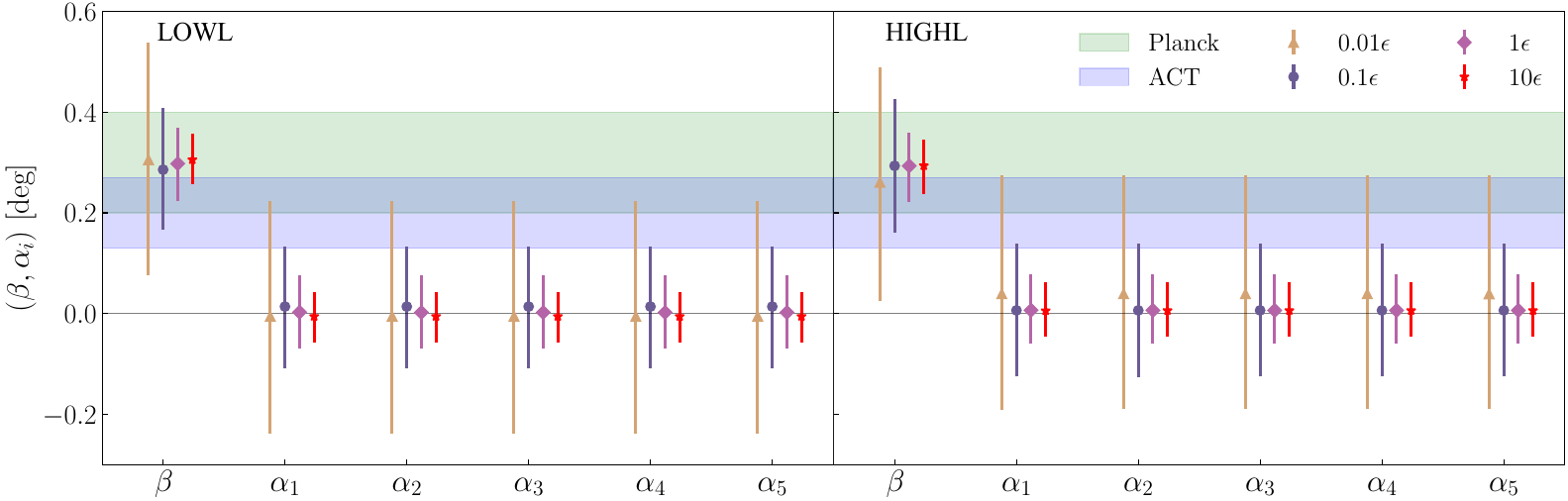}
    \caption{Inferred isotropic rotation angles \{$\beta$, $\alpha_i$\} from the $EE$, $BB$ and $EB$ power spectra. We consider three different levels of foreground polarization for the \textbf{LOWL} case (left) and the \textbf{HIGHL} case (right). The shaded regions correspond to the Planck~\citep{mk} and ACT~\citep{actcb} constraints on $\beta$.}
    \label{fig:error-bar}
\end{figure*}

\begin{figure*}[ht]
    \centering
    \includegraphics[width=1\linewidth]{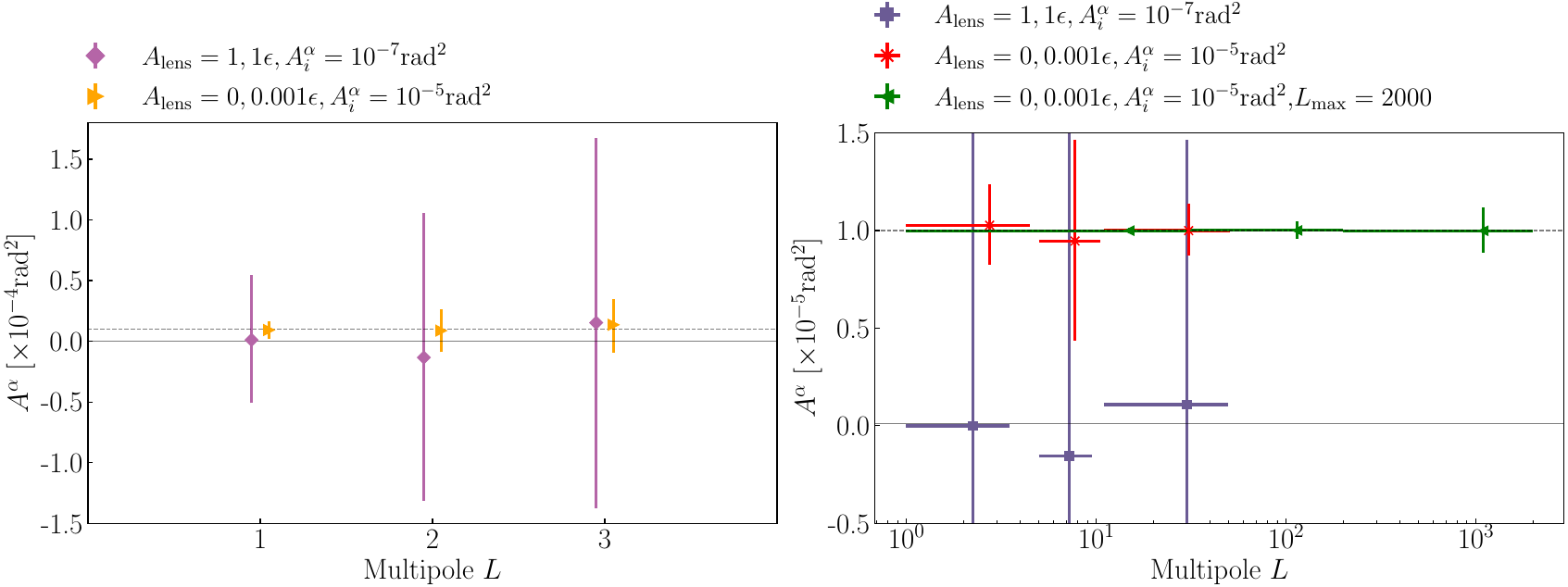}
    \caption{Reconstructed power spectra of the anisotropic cosmic birefringence from the $EE$, $BB$ and $EB$ power spectra for the $\textbf{LOWL}$ and the $\textbf{HIGHL}$ cases. We investigate how the reconstruction of different input signals can be affected by both the lensing and the foreground signals.}
    \label{fig:claa}
\end{figure*}

\begin{figure}[ht]
    \centering
    \includegraphics[width=1\linewidth]{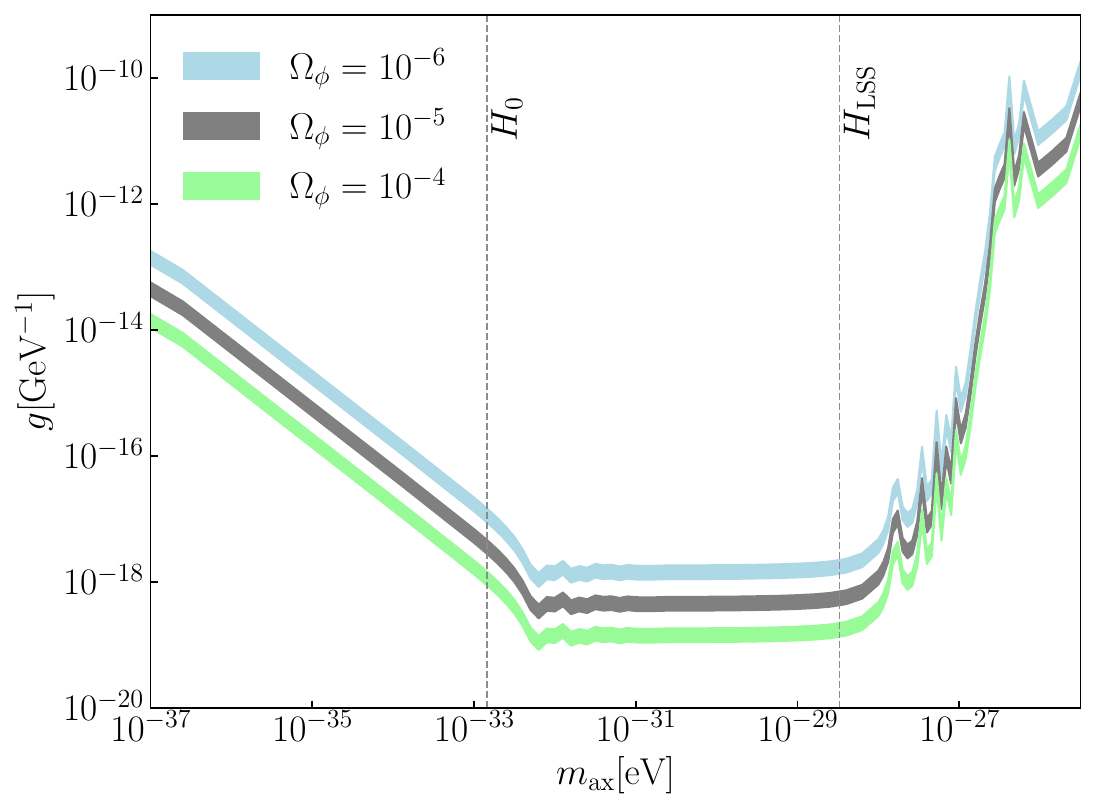}
    \caption{The constraints on the ALP-photon coupling constant $g$ with the inferred isotropic angles. We assume a quadratic potential for the ALPs, i.e., $V=1/2(m_{\rm ax}\phi)^2$. Here $\phi$ refers to the pseudoscalar field of ALP and $m_{\rm ax}$ denotes its mass. We generate the mock data with a fiducial parameter set $(A_{\rm lens}, \beta)=(1.0, 0.3^{\circ})$ for the \textbf{LOWL} at the 1$\epsilon$ foreground level, and obtain the posterior distributions of the isotropic angles. We can constrain the ALP-photon coupling constant in a broad range of ALP energy density fractions ($\Omega_{\phi}$)~\cite{PhysRevD.103.043509}. In this plot, $H_0$ and $H_{\rm LSS}$ refer to the Hubble parameters today and at the last scattering surface, respectively.}
    \label{fig:axion}
\end{figure}

\section{Inference of different rotation angles with a Bayesian analysis}
\label{sec:method}

In the previous section, we discussed how to generate mock power spectra with the fiducial parameter set in Table~\ref{table:priors}. In this section, we investigate how to infer different parameters using a Bayesian analysis. We assume uniform priors for the parameters and establish a likelihood function as
\begin{equation}
    -2\ln\mathcal{L} = \sum_\ell\left[\boldsymbol{\hat x}_\ell(\Theta^\ast) -\boldsymbol{x}_\ell(\Theta)\right]^{T} 
    {\mathcal{C}_\ell}^{-1} \left[\boldsymbol{\hat x}_\ell(\Theta^\ast) -\boldsymbol{x}_\ell(\Theta)\right],
\end{equation}
where $\boldsymbol{x}_\ell(\Theta)$ is a data vector constructed from $EE$, $BB$ and $EB$ power spectra, $\boldsymbol{\hat x}_\ell(\Theta^\ast)$ refers to the mock data vector with the fiducial parameter set $\Theta^\ast$ and the superscript ``$T$" denotes a matrix transpose operation. $\mathcal{C}_\ell$ is the covariance matrix calculated at each multipole $\ell$. The covariance matrix has a dimension of $2n_\nu^2 + n_\nu$ with $n_{\nu}$ frequency bands, so the dimension is 55 for five bands considered in this work. We calculate each covariance matrix $\mathcal{C}_\ell$ from the definition 
\begin{equation}
    \mathcal{C}_\ell = \mathrm{Cov}\bigl(\hat{\boldsymbol{x}}_\ell(\Theta^\ast), \hat{\boldsymbol{x}}^T_\ell(\Theta^\ast)\bigr),
\end{equation}
which can be further simplified as 
\begin{equation}
    \begin{split}
        \mathcal{C}_\ell({\hat{\boldsymbol{x}}}^{a\times b}_{\ell}, {\hat{\boldsymbol{x}}}^{c\times d}_{\ell}) 
            &= \frac{1}{(2\ell+1)f_{\rm sky}\Delta\ell} \left(\hat{C}^{a\times c}_{\ell}\hat{C}^{b\times d}_{\ell}\right. \\
               &\quad\ \left. + \hat{C}^{a\times d}_{\ell}\hat{C}^{b\times c}_{\ell}\right).
    \end{split}
    \label{matrix_element_covariance}
\end{equation}
Here, we consider a full sky coverage and a multipole range $2 \leq \ell \leq 1000$ with $\Delta\ell=1$. To improve the numerical stability of the covariance matrix, we adopt the same approach as used in~\cite{2020PTEP.2020j3E02M} where the off-diagonal block matrices such as $EEEB$ and $EEBB$ are ignored to ensure the numerical stability of the matrix inversion. We have tested that neglecting these terms can make the constraints less stringent so this approach is conservative. We sample the posterior probability distribution functions (PDFs) using the Markov chain Monte Carlo (MCMC) method via the package \texttt{emcee}~\cite{2013PASP..125..306F}, with which the MCMC convergence is assessed using the autocorrelation time. The parameter priors are assumed to be uniform and are listed in Table~\ref{table:priors}. 

We also employ the Fisher Matrix (FM) formalism to cross-check the MCMC results. The FM is a useful tool for exploring the posterior parameter space of a nearly Gaussian model and is essentially a second-order derivative of the log-likelihood function:
\begin{equation}
    F_{ij} \equiv \left.\Big\langle-\frac{\partial^2\ln\mathcal{L}}{\partial\Theta_i\partial\Theta_j}\Big\rangle\right|_{\Theta=\Theta^*}, 
    \label{eqn:fisher-matrix}
\end{equation}
which can be numerically evaluated by the second-order central-difference form 
\begin{eqnarray}
        \frac{\partial^2f(x, y)}{\partial x\partial y} 
          &\simeq& \frac{1}{4\Delta x\Delta y} [f(x + \Delta x, y + \Delta y) \nonumber\\
          &&- f(x + \Delta x, y - \Delta y)\nonumber\\
          &&- f(x - \Delta x, y + \Delta y) \nonumber\\
          &&+ f(x - \Delta x, y - \Delta y)].
    \label{eqn:center-difference}
\end{eqnarray}
We choose the step sizes $\Delta x$ and $\Delta y$ carefully to ensure numerical convergence.

\section{Results and discussions}
\label{sec:result}

We simulate the mock data with different components depicted in Fig.~\ref{fig:power-all} and also include the instrumental noise and beam profiles. We adopt the Bayesian approach as described in Sec.~\ref{sec:method} to infer different parameters from the multifrequency data which are constructed by power-spectrum values at different frequency pairs and multipoles. 
The parameters defined in Table~\ref{table:priors} consist of cosmological parameters $A_{\mathrm{lens}}$, $\beta$, $\{A^{\alpha}_{k}\}$, miscalibration angles $\{\alpha_i\}$ and foreground parameters $A_c$, $\alpha_c$, $B_c$ where $c=D,S$. We show the PDFs for three types of parameters in Figs.~\ref{fig:posterior1}-\ref{fig:posterior6}, which are extracted from the full parameter contours. The inferred parameters are listed in Table \ref{table:priors}.

\begin{table}[htbp]
\centering
\caption{We assume uniform priors for different parameters. The isotropic rotation angle is taken from the recent Planck measurement~\cite{mk}. We also assume that the anisotropic rotation signals are much lower than the current limits~\cite{spt,actcb,bkcb}. We show the estimated parameters for the \textbf{LOWL} case at a 1$\epsilon$ foreground level.}
\label{table:priors}
\begin{tabular}{cccc}
\toprule
\textbf{Parameters}       & \textbf{Priors}  & \textbf{Fiducial} & \textbf{MCMC} \\
\midrule
$A_{\mathrm{lens}}$        & $[0, 2]$        & $1$              & $1.00\pm 4\times 10^{-3}$ \\
$\beta (^\circ)$      &  $[-1, 1]$      & $0.3$            & $0.30\pm0.07$ \\
$A^{\alpha}_1$(rad$^2$)    & $[-0.001, 0.001]$ & $10^{-7}$ & $(0.002\pm0.5)\times 10^{-4}$ \\ 
$A^{\alpha}_2$(rad$^2$)    & $[-0.001, 0.001]$ & $10^{-7}$ & $(-0.1\pm1.2)\times 10^{-4}$ \\ 
$A^{\alpha}_3$(rad$^2$)    & $[-0.001, 0.001]$ & $10^{-7}$ & $(0.2\pm 1.5)\times 10^{-4}$ \\ 
$\alpha_1 (^\circ)$   & $[-1, 1]$       & $0$              & $0.001^{+0.075}_{-0.071}$ \\
$\alpha_2 (^\circ)$   & $[-1, 1]$       & $0$              & $0.001^{+0.075}_{-0.071}$\\
$\alpha_3 (^\circ)$   & $[-1, 1]$       & $0$              & $0.0009^{+0.075}_{-0.071}$ \\
$\alpha_4 (^\circ)$   & $[-1, 1]$       & $0$              & $0.001^{+0.075}_{-0.071}$ \\
$\alpha_5 (^\circ)$   & $[-1, 1]$       & $0$              & $0.0009^{+0.075}_{-0.071}$ \\
$A_D (\mu\mathrm{K}^2)$  & $[0, 50]$    & $28$             & $28.0\pm 0.1$ \\
$\alpha_D$            & $[-0.2, 0]$     & $-0.16$          & $-0.16\pm 0.002$ \\
$B_D$                 & $[-10^{-6}, 10^{-6}]$ & $8\times10^{-9}$ & $(8\pm1.3)\cdot10^{-9}$ \\
$A_S (\mu\mathrm{K}^2)$  & $[0, 5]$     & $1.6$            & $1.60^{+0.009}_{-0.008}$ \\
$\alpha_S$            & $[-1, 0]$       & $-0.93$          & $-0.93\pm 0.004$ \\
$B_S$                 & $[-10^{-6}, 10^{-6}]$ & $3\times10^{-9}$ & $(3.07^{+0.9}_{-1.0})\cdot10^{-9}$ \\
\bottomrule
\end{tabular}
\end{table}

In Fig.~\ref{fig:posterior1}, we show the PDFs of the cosmological parameters for the \textbf{LOWL} and the \textbf{HIGHL} scenarios. The isotropic rotation angle is negligibly correlated with either the lensing or the anisotropic components, and the constraints on the isotropic rotation angles are also weakened by decreasing the amplitudes of foreground polarization. This is consistent with the MK predictions and indicates that foreground-mitigated polarization data may not have enough sensitivity to constrain the isotropic angles. However, we can address this issue in two aspects. Experimentally, we can set stringent priors on the miscalibration angles using the optics models~\cite{actcb} and high-precision calibration sources~\cite{bkiso}. Theoretically, we can incorporate the ALP dynamics to break the degeneracies between the two types of isotropic angles~\cite{alpmasses}. Both the gravitational lensing and anisotropic cosmic birefringence introduce small-scale perturbations to the CMB polarization power spectra, but there are only small correlations among the contours for lensing and $A_i^{\alpha}$ due to the limited $L_{\mathrm{max}}$ considered in both the \textbf{LOWL} and the \textbf{HIGHL} scenarios. Interestingly, there are correlations among different $A_i^{\alpha}$s and these correlations may help differentiate the massless ALPs from the massive ones, corresponding to scale-invariant and scale-dependent rotation power spectra, respectively~\cite{alpmasses}. 

We also adopt the Fisher Matrix formalism to predict the parameter constraints in Fig.~\ref{fig:fisher} as a sanity check. The contours from the Fisher Matrix calculations are consistent with the MCMC results in Fig.~\ref{fig:posterior1}. We adopt the MCMC method as the baseline approach in this work. 

In Fig.~\ref{fig:posterior1}, we notice that the constraints of the isotropic rotation angles are mainly determined by the foreground polarization amplitudes, while other foreground parameters such as the spectral indices $\alpha_c$ and higher-order fluctuations $B_c$ have a very negligible impact, as demonstrated in Fig.~\ref{fig:posterior5}. This indicates that the spatially-varying SEDs may not affect the isotropic angle detection, and the standard foreground removal procedures, such as the internal linear combinations (ILC) and needlet ILC, can be sufficient for the foreground removal task of the rotation angle analyses.

In addition to the foreground removal, a delensing procedure can also be applied to the CMB observations. For a future polarization experiment with an ultra-low noise level below 5$\mu {\rm K}{\mbox{-}}{\rm arcmin}$, the lensing contribution is more significant than the instrumental noise. Instead of performing a map-level delensing procedure, which is beyond the scope of this work, we adopt a single parameter $A_{\rm lens}$ to quantify the residual lensing signals at the power-spectrum level and assess the impact of the lensing contributions in Fig.~\ref{fig:posterior-beta_A_CB1_A_CB2_A_CB3_BOTH_LOWL_HIGHL_Alens1_Alens0}. It is seen that the constraints on the anisotropic rotation angles can be improved once the delensing procedures are incorporated and this is consistent with the results from the quadratic-estimator-based delensing~\cite{lonappan25}. Similarly, we could also expect to gain better constraints on the isotropic angles if there are no instrumental miscalibration effects~\cite{lonappan25}. However, in this work we consider the polarization signals generated by the miscalibration angles which are degenerate with the isotropic ones, and thus the improvements on the isotropic angles are compromised.

The miscalibration angles are strongly correlated among different frequency bands but can be constrained to $\sim0.07^{\circ}$ for the low-noise CMB experiment with a $1\epsilon$-level foreground contamination considered in this work. Fig.~\ref{fig:posterior6} shows that these angles have very mild correlations with the foreground parameters except the foreground amplitudes $A_D$ and $A_S$. 
We combine the constraints on different rotation angles in Fig.~\ref{fig:error-bar} for the \textbf{LOWL} and the \textbf{HIGHL} cases in the left and right subpanels, respectively. Overall, the foreground polarization amplitudes can significantly affect the constraints on the isotropic rotation angles, which are negligibly affected by the anisotropic rotation angles as indicated by both the \textbf{LOWL} and the \textbf{HIGHL} cases.

We show the reconstructed power spectra of the anisotropic rotation angles with different mock datasets in Fig.~\ref{fig:claa}. For the mock data with a 0.001$\epsilon$-level foreground polarization and a full delensing, we validate that the pipeline can precisely extract the anisotropic rotation power spectrum with an amplitude of $A^{\alpha}_i=10^{-5}{\rm rad^2}$ for both the \textbf{LOWL} and the \textbf{HIGHL} cases, although such an amplitude has been ruled out by recent CMB polarization measurements~\cite{spt, actcb2, bkcb}. However, we find it would be challenging to achieve such a detection from the CMB polarization power spectra alone if the global amplitude is $A^{\alpha}_i=10^{-7}{\rm rad^2}$. Nevertheless, the reconstructed piece-wise rotation power spectra in Fig.~\ref{fig:claa} would be a complementary probe to the ones generated from the four-point correlation functions with the quadratic estimators since they provide the $\mathcal{O}(\delta\alpha^2)$ level fluctuations whereas the quadratic estimators can mainly extract the $\mathcal{O}(\delta\alpha)$ level fluctuations. 

We can use the inferred isotropic rotation angles in Fig.~\ref{fig:error-bar} to study the exotic physics of axion-like particles (ALPs). By assuming a 0.3-degree cosmological rotation angle under different energy density fractions $\Omega_a$ of the ALPs, we can place stringent limits on the ALP-photon coupling coefficient $g$ of the Chern-Simons coupling term $g\phi F\tilde F$ in a broad range of ALP masses. Here, $\phi$ and $F$ denote the pseudoscalar and electromagnetic fields, respectively. In this work, we consider a quadratic ALP potential $V=1/2(m_{\rm ax}\phi)^2$ where $m_{\rm ax}$ denotes the ALP mass. We solve the background evolution of the pseudoscalar field at a given fraction $\Omega_a$ which determines the initial field value and derive the $g\mbox{-}m_{\rm ax}$ relation according to the expression of the ALP-induced rotation angle $\beta=g\Delta\phi/2$. In Fig.~\ref{fig:axion}, we show the 1-$\sigma$ uncertainty ranges for the $g\mbox{-}m_{\rm ax}$ predictions at different ALP energy density fractions. The most sensitive mass range for the isotropic angles is within $10^{-33}{\rm eV}<m_{\rm ax}<10^{-28}{\rm eV}$ corresponding to the epochs of today and the last scattering surface.

\section{Conclusions}
\label{sec:conclusion}

In this work, we establish a new formalism to simultaneously infer both the isotropic and anisotropic rotation angles. For the next-generation high-sensitivity CMB polarization experiments, this new formalism can be applied to foreground-suppressed polarization data taken at multiple frequencies.

We model the polarization power spectra for both the three rotation effects and the foreground contaminants with higher-order spatial variations. We generate mock power spectra at the nominal CMB frequencies assuming futuristic instrumental settings. With the likelihood function built from the power-spectrum model, we adopt the Bayesian analysis to infer the cosmological rotation angles, miscalibration angles and the foreground parameters using different mock data with varied foreground levels, multipole ranges and lensing contamination. We validate that the formalism presented in this work can successfully extract the detailed information of both the cosmological signals and polarized foregrounds from the multifrequency power spectra. Also, we find that the inferred posterior distributions of different parameters are consistent with the fiducial model. Especially, we explore interesting correlations among different parameters which can serve as useful guidance for achieving better constraints for the future datasets.

The reconstruction of anisotropic cosmic birefringence from this approach is complementary to the four-point correlation functions constructed by the quadratic estimators~\cite{4ptvera}. Both approaches can be applied to the future CMB datasets to detect the spatially varying rotation effects, which will be crucial for studying the physics of axion-like particles~\cite{komatsu} and unveiling the nature of dynamical dark energy~\cite{alpde}.

\acknowledgments
We are grateful for the helpful discussions with Christian Reichardt, Wen Zhao, and Mingzhe Li. C. F. acknowledges the support from USTC and F. B. A acknowledges the support from CAS.

\bibliography{bibliography}

\clearpage
\onecolumngrid
\appendix
\section{CMB power spectra perturbations by anisotropic cosmic rotation effects using correlation functions}
\label{appsec:derivation-cb}

We can define three correlation functions for the CMB polarization vectors $P=Q\pm \mathrm{i}U$ as
\begin{align}
    \tilde{\xi}_{+} 
      &\equiv \left<[\tilde{Q} + {\rm i}\tilde{U}]^*(\hat{n}) [\tilde{Q} + {\rm i}\tilde{U}](\hat{n^\prime})\right>,
      \label{eqn:xip} \\
    {\xi}_{+}^{ij} 
      &\equiv \left<[Q_i + {\rm i}U_i]^*(\hat{n}) [Q_j + {\rm i}U_j](\hat{n^\prime})\right>,
      \label{eqn:xipij} \\
    {\xi}_{+}^{ji} 
      &\equiv \left<[Q_j + {\rm i}U_j]^*(\hat{n}) [Q_i + {\rm i}U_i](\hat{n^\prime})\right>.
      \label{eqn:xipji}
\end{align}
Here, $\tilde Q$ and $Q$ (or $\tilde U$ and $U$) denote the Stokes parameters without and with rotation effects. The indices $i, j$ refer to frequency pairs and $\hat n$ denotes a sky direction. 

We define an angle $\theta$ between two directions with $\cos\theta = \hat{n}\cdot\hat{n^\prime}$ and can use the mathematical properties of the Wigner d-function $d_{22}^\ell$ to simplify the correlation function $\tilde{\xi}_{+}$. Similar to the methods introduced in~\cite{corr1, corr2, Mingzhe_Li_2013}, we can substitute the CMB power spectra into the correlation functions so that they can be expressed as
\begin{equation}
    \begin{split}
        \tilde{\xi}_{+}(\theta)
          &= \sum_\ell\frac{2\ell + 1}{4\pi}(\tilde{C}^{{E}{E}}_\ell + \tilde{C}^{{B}{B}}_{\ell})d^{\ell}_{22}(\theta).
    \end{split}
    \label{eqn:xi-plus}
\end{equation}

Similarly, we can substitute the spherical harmonic expansion of CMB polarization into Eq.~(\ref{eqn:xipij}) and simplify the correlation functions ${\xi}_{+}^{ij}$ with the CMB power spectra
\begin{equation}
    \begin{split}
        \xi_{+}^{ij}(\theta)
          &= \sum_{\ell m, \ell^\prime m^\prime} 
             \left<({E^i_{\ell m}}^* - {\rm i}{B^i_{\ell m}}^*)(E_{\ell^\prime m^\prime}^j + {\rm i}B_{\ell^\prime m^\prime}^j)\right> {}_{2}Y^{*} _{\ell m}(\hat{n})_{2}Y_{\ell^{\prime}m^{\prime}}(\hat{n^{\prime}})\\
          &= \sum_{\ell m}({C}^{rot, {EE}, (ij)}_{\ell} + {C}^{rot, {BB}, (ij)}_{\ell} + {\rm i}C_\ell^{rot, {EB}, (ij)} - {\rm i}C_\ell^{rot, {EB}, (ji)}) {}_{2}Y ^{*}_{\ell m}(\hat{n})_{2}Y_{\ell m}(\hat{n^{\prime}})\\
          &= \sum_{\ell}\frac{2\ell+1}{4\pi}({C}^{rot, {EE}, (ij)}_{\ell} + {C}^{rot, {BB}, (ij)}_{\ell} + {\rm i}C_\ell^{rot, {EB}, (ij)} - {\rm i}C_\ell^{rot, {EB}, (ji)}) d^{\ell}_{22}(\theta),
    \end{split}
    \label{eqn:xi-plus-ij}
\end{equation}
where we have used the relation between the spin-weighted spherical harmonics and Wigner d-function 
\begin{equation}
    \sum_{m=-\ell}^{\ell}{}_2Y_{\ell m}^*(\hat{n}) {}_2Y_{\ell m}(\hat{n}^\prime)  = \frac{2\ell + 1}{4\pi} d_{22}^\ell(\theta).
\end{equation}
Similarly, we can express the correlation function $\xi_{+}^{ji}(\theta)$ as
\begin{equation}
    \begin{split}
        \xi_{+}^{ji}(\theta)
          &= \sum_{\ell}\frac{2\ell+1}{4\pi}({C}^{rot, {EE}, (ji)}_{\ell} + {C}^{rot, {BB}, (ji)}_{\ell} + {\rm i}C_\ell^{rot, {EB}, (ji)} - {\rm i}C_\ell^{rot, {EB}, (ij)}) d^{\ell}_{22}(\theta).
    \end{split}
    \label{eqn:xi-plus-ji}
\end{equation}

Given the fact that $C_\ell^{{EE}, ij} = C_\ell^{{EE}, ji}$ and $C_\ell^{{BB}, ij} = C_\ell^{{BB}, ji}$, but $C_\ell^{{EB}, ij} \neq C_\ell^{{EB}, ji}$, the correlation functions are not symmetric, i.e., $\xi_+^{ij}(\theta) \neq \xi_+^{ji}(\theta)$. Using the orthogonal property of the Wigner d-function, we can integrate the correlation function in Eq.~(\ref{eqn:xi-plus-ij}) with the d-function as 
\begin{equation}
    \begin{split}
        \int_{-1}^1 \xi_+^{ij}(\theta) d_{22}^\ell(\theta) \mathrm{d}\cos\theta
          &= \sum_{\ell^\prime}\frac{2\ell^\prime + 1}{4\pi}({C}^{rot, {EE}, (ij)}_{\ell^\prime} + {C}^{rot, {BB}, (ij)}_{\ell^\prime} + {\rm i}C_{\ell^\prime}^{rot, {EB}, (ij)} - {\rm i}C_{\ell^\prime}^{rot, {EB}, (ji)})
             \int_{-1}^1 d_{22}^\ell(\theta)d^{\ell^\prime}_{22}(\theta) \mathrm{d}\cos\theta \\
          &= \frac{1}{2\pi}({C}^{rot, {EE}, (ij)}_{\ell} + {C}^{rot, {BB}, (ij)}_{\ell} + {\rm i}C_{\ell}^{rot, {EB}, (ij)} - {\rm i}C_{\ell}^{rot, {EB}, (ji)}).
    \end{split}
\end{equation}
The expression above implies that
\begin{equation}
    \begin{split}
        &{C}^{rot, {EE}, (ij)}_{\ell} + {C}^{rot, {BB}, (ij)}_{\ell} + {\rm i}C_{\ell}^{rot, {EB}, (ij)} - {\rm i}C_{\ell}^{rot, {EB}, (ji)} = 2\pi \int_{-1}^1 \xi_+^{ij}(\theta) d_{22}^\ell(\theta) \mathrm{d}\cos\theta,
    \end{split}
     \label{eqn:xi-plus-ij-reverse}
\end{equation}
and 
\begin{equation}
    \begin{split}
        &{C}^{rot, {EE}, (ji)}_{\ell} + {C}^{rot, {BB}, (ji)}_{\ell} + {\rm i}C_{\ell}^{rot, {EB}, (ji)} - {\rm i}C_{\ell}^{rot, {EB}, (ij)} = 2\pi \int_{-1}^1 \xi_+^{ji}(\theta) d_{22}^\ell(\theta) \mathrm{d}\cos\theta.
    \end{split}
     \label{eqn:xi-plus-ji-reverse}
\end{equation}

By substituting Eq.~(\ref{eqn:rotated-stokes}) into Eq.~(\ref{eqn:xipij}), we can obtain
\begin{equation}
    \begin{split}
        {\xi}_{+}^{ij}(\theta) 
          &= \exp[2{\rm i}(\alpha_j - \alpha_i)] 
             \exp\left[4C^\alpha(\theta) - 4C^\alpha(0)\right]
             \tilde{\xi}_+(\theta).
    \end{split}
    \label{eqn:xi-plus-ij-angle}
\end{equation}
where we have used the relation $\left<\exp(\mathrm{i}x)\right> = \exp(-\left<x^2\right>/2)$ for an arbitrary Gaussian variable $x$ with a zero mean. 

Combining Eq.~(\ref{eqn:xi-plus-ij-reverse}) and Eq.~(\ref{eqn:xi-plus-ij-angle}), we can establish the relations between the unrotated and rotated CMB power spectra as follows
\begin{equation}
    \begin{split}
        {C}^{rot, {EE}, (ij)}_{\ell} + {C}^{rot, {BB}, (ij)}_{\ell} + {\rm i}C_{\ell}^{rot, {EB}, (ij)} - {\rm i}C_{\ell}^{{rot, EB}, (ji)} &= e^{2{\rm i}(\alpha_j - \alpha_i)} e^{-4C^\alpha(0)}
           \sum_{\ell^\prime}
           \frac{2\ell^\prime + 1}{2}(\tilde{C}^{{E}{E}}_{\ell^\prime} + \tilde{C}^{{B}{B}}_{\ell^\prime}) \\
        &\quad \times
           \int_{-1}^1 
           d_{22}^{\ell^\prime}(\theta) d_{22}^{\ell}(\theta) e^{4C^\alpha(\theta)} \mathrm{d}\cos\theta.
    \end{split}
    \label{eqn:final-eeij-plus-bbij-plus-ebij-minus-ebji}
\end{equation}

By exchanging the indices $i$ and $j$, we also get 
\begin{equation}
    \begin{split}
        {C}^{rot, {EE}, (ji)}_{\ell} + {C}^{rot, {BB}, (ji)}_{\ell} + {\rm i}C_{\ell}^{rot, {EB}, (ji)} - {\rm i}C_{\ell}^{rot, {EB}, (ij)} &=e^{2{\rm i}(\alpha_i - \alpha_j)} e^{-4C^\alpha(0)}
           \sum_{\ell^\prime}
           \frac{2\ell^\prime + 1}{2}(\tilde{C}^{{E}{E}}_{\ell^\prime} + \tilde{C}^{{B}{B}}_{\ell^\prime}) \\
        &\quad \times
           \int_{-1}^1 
           d_{22}^{\ell^\prime}(\theta) d_{22}^{\ell}(\theta) e^{4C^\alpha(\theta)} \mathrm{d}\cos\theta.
    \end{split}
    \label{eqn:final-eeji-plus-bbji-plus-ebji-minus-ebij}
\end{equation}

Furthermore, we define other correlation functions as
\begin{align}
    \tilde{\xi}_{-} 
      &\equiv \left<[\tilde{Q} + {\rm i}\tilde{U}](\hat{n}) [\tilde{Q} + {\rm i}\tilde{U}](\hat{n^\prime})\right>,
      \label{eqn:xim} \\
    {\xi}_{-}^{ij} 
      &\equiv \left<[Q_i + {\rm i}U_i](\hat{n}) [Q_j + {\rm i}U_j](\hat{n^\prime})\right>,
      \label{eqn:ximij} \\
    {\xi}_{-}^{ji} 
      &\equiv \left<[Q_j + {\rm i}U_j](\hat{n}) [Q_i + {\rm i}U_i](\hat{n^\prime})\right>.
      \label{eqn:ximji}
\end{align}

Repeating the derivations above, we obtain
\begin{equation}
    \begin{split}
        {C}^{rot, {EE}, (ij)}_{\ell} - {C}^{rot, {BB}, (ij)}_{\ell} + {\rm i}C_{\ell}^{rot, {EB}, (ij)} + {\rm i}C_{\ell}^{rot, {EB}, (ji)} &= e^{{\rm i}(4\beta + 2\alpha_i + 2\alpha_j)} e^{-4C^\alpha(0)}
           \sum_{\ell^\prime}
           \frac{2\ell^\prime + 1}{2}(\tilde{C}^{{E}{E}}_{\ell^\prime} - \tilde{C}^{{B}{B}}_{\ell^\prime} + 2{\rm i}\tilde{C}_{\ell^\prime}^{{E}{B}}) \\
        &\quad \times
           \int_{-1}^1 
           d_{-22}^{\ell^\prime}(\theta) d_{-22}^{\ell}(\theta) e^{-4C^\alpha(\theta)} \mathrm{d}\cos\theta.
    \end{split}
    \label{eqn:final-eeij-minus-bbij-plus-ebij-plus-ebji}
\end{equation}

From Eq.~(\ref{eqn:final-eeij-plus-bbij-plus-ebij-minus-ebji}) and Eq.~(\ref{eqn:final-eeji-plus-bbji-plus-ebji-minus-ebij}), we can solve the terms ${C}^{rot, {EE}, (ij)}_{\ell} + {C}^{rot, {BB}, (ij)}_{\ell}$ and ${C_{\ell}^{rot, {EB}, (ij)} - C_{\ell}^{rot, {EB}, (ji)}}$ as
\begin{align}
    {C}_{\ell}^{rot, {{EE}}, (ij)} + {C}_{\ell}^{rot, {BB}, (ij)}
      &= {\cos(2\alpha_i - 2\alpha_j)}
        e^{-4C^{\alpha}(0)} 
        \sum_{\ell^{\prime}} \frac{2\ell^{\prime} + 1}{2} 
        (\tilde{C}_{\ell^{\prime}}^{{E}{E}} + \tilde{C}_{\ell^{\prime}}^{{B}{B}})
        \int_{-1}^{1} d_{22}^{\ell^{\prime}}(\theta) d_{22}^{\ell}(\theta) 
        e^{4C^{\alpha}(\theta)} \mathrm{d}\cos\theta, 
        \label{eqn:final-eeij-plus-bbij} \\
    {{C}_{\ell}^{rot, {EB}, (ij)} - C_\ell^{rot, {EB}, (ji)}}
	  &{= \sin(2\alpha_j - 2\alpha_i) e^{-4C^{\alpha}(0)}
	     \sum_{\ell^{\prime}}\frac{2\ell^{\prime} + 1}{2}
	     (\tilde{C}_{\ell^{\prime}}^{EE} + \tilde{C}_{\ell^{\prime}}^{BB})
	     \int_{-1}^{1} d_{22}^{\ell^{\prime}}(\theta) d_{22}^{\ell}(\theta)
	     e^{4C^{\alpha}(\theta)}\mathrm{d}\cos\theta.
	     \label{eqn:final-ebij-minus-ebji}}
\end{align}

Given the fact that the Wigner d-function and the CMB power spectra are both real functions, we can split the right side of Eq.~(\ref{eqn:final-eeij-minus-bbij-plus-ebij-plus-ebji}) into the real and imaginary parts and get\footnote{Note we have assumed that the CMB primordial $EB$ correlation is vanishing, i.e., $\tilde{C}_\ell^{{E}{B}} = 0$.}

\begin{align}
    {C}_{\ell}^{rot, {EE}, (ij)} - {C}_{\ell}^{rot, {BB}, (ij)} 
	  &= \cos(4\beta {+ 2\alpha_i + 2\alpha_j}) e^{-4C^{\alpha}(0)}
	     \sum_{\ell^{\prime}}\frac{2\ell^{\prime} + 1}{2}
	     (\tilde{C}_{\ell^{\prime}}^{{E}{E}} - \tilde{C}_{\ell^{\prime}}^{{B}{B}})
	     \int_{-1}^{1} d_{-22}^{\ell^{\prime}}(\theta) d_{-22}^{\ell}(\theta)
	     e^{-4C^{\alpha}(\theta)}\mathrm{d}\cos\theta, 
	     \label{eqn:final-eeij-minus-bbij} \\
    {C}_{\ell}^{rot, {EB}, (ij)} + C_\ell^{rot, {EB}, (ji)}
	  &= \sin(4\beta {+ 2\alpha_i + 2\alpha_j}) e^{-4C^{\alpha}(0)}
	     \sum_{\ell^{\prime}}\frac{2\ell^{\prime} + 1}{2}
	     (\tilde{C}_{\ell^{\prime}}^{{E}{E}} - \tilde{C}_{\ell^{\prime}}^{{B}{B}})
	     \int_{-1}^{1} d_{-22}^{\ell^{\prime}}(\theta) d_{-22}^{\ell}(\theta)
	     e^{-4C^{\alpha}(\theta)}\mathrm{d}\cos\theta.
	     \label{eqn:final-ebij-plus-ebji}
\end{align}

We have cross-checked the derivations with no miscalibration angles, i.e., $\alpha_i = 0$ for all frequency bands $i$, and the rotated power spectra are consistent with the ones introduced in~\cite{Mingzhe_Li_2013}.

The integrations with the Wigner d-functions are very time-consuming. Therefore, we Taylor expand the exponential terms involved in Eqs.~(\ref{eqn:final-eeij-plus-bbij})-(\ref{eqn:final-ebij-plus-ebji}) since we assume that the anisotropic rotation effects are secondary, so $C^\alpha(\theta) < C^\alpha(0) \ll 1$. We only keep the first order terms and ignore other higher order terms. For example, we can simplify the integral in Eq.~(\ref{eqn:final-eeij-plus-bbij}) as
\begin{equation}
    \begin{split}
        \int_{-1}^1 d_{22}^{\ell^\prime}(\theta) d_{22}^\ell(\theta) e^{4C^\alpha(\theta)} \mathrm{d}\cos\theta 
            &\simeq \int_{-1}^1 d_{22}^{\ell^\prime}(\theta) d_{22}^\ell(\theta) \left[1 + 4C^\alpha(\theta)\right] \mathrm{d}\cos\theta \\
            &= \int_{-1}^1 d_{22}^{\ell^\prime}(\theta) d_{22}^\ell(\theta) \mathrm{d}\cos\theta +
               \int_{-1}^1 d_{22}^{\ell^\prime}(\theta) d_{22}^\ell(\theta) \sum_L\frac{2L + 1}{\pi}C_L^{\alpha\alpha}P_L(\cos\theta) \mathrm{d}\cos\theta \\
            &= \frac{2}{2\ell + 1}\delta_{\ell\ell^\prime} + 
               2\sum_L \frac{2L + 1}{\pi}C_L^{\alpha\alpha} 
               \begin{pmatrix}
               \ell & L & \ell^\prime \\
               2    & 0 & -2
               \end{pmatrix}^2,
    \end{split}
    \label{eqn:integral-22}
\end{equation}
where we have used the properties of the Wigner d-functions. With these simplifications, we can express the linear combinations of the rotated power spectra as
\begin{align}
    {C}_{\ell}^{rot, {EE}, (ij)} + {C}_{\ell}^{rot, {BB}, (ij)}
      &\simeq 
        \cos(2\alpha_i - 2\alpha_j)
        e^{-4C^{\alpha}(0)} 
        \biggl[(\tilde{C}_{\ell}^{{E}{E}} + \tilde{C}_{\ell}^{{B}{B}})
        \nonumber\\
      &\quad +
        \sum_{\ell^\prime L} \frac{(2\ell^\prime + 1)(2L + 1)}{\pi} C_L^{\alpha\alpha}
        \begin{pmatrix}
            \ell & L & \ell^\prime \\
            2    & 0 & -2
        \end{pmatrix}^2
        (\tilde{C}_{\ell^{\prime}}^{{E}{E}} + \tilde{C}_{\ell^{\prime}}^{{B}{B}})\biggr],
      \label{eqn:final-eeij-plus-bbij-2} \\
    {C}_{\ell}^{rot, {EE}, (ij)} - {C}_{\ell}^{rot, {BB}, (ij)}
      &\simeq 
        \cos(4\beta + 2\alpha_i + 2\alpha_j)\ e^{-4C^{\alpha}(0)} 
        \biggl[(\tilde{C}_{\ell}^{{E}{E}} - \tilde{C}_{\ell}^{{B}{B}})
        \nonumber\\
      &\quad -
        \sum_{\ell^\prime L} \frac{(2\ell^\prime + 1)(2L + 1)}{\pi} C_L^{\alpha\alpha}
        (-1)^{\ell + L + \ell^\prime}
        \begin{pmatrix}
            \ell & L & \ell^\prime \\
            2    & 0 & -2
        \end{pmatrix}^2
        (\tilde{C}_{\ell^{\prime}}^{{E}{E}} - \tilde{C}_{\ell^{\prime}}^{{B}{B}})\biggr],
      \label{eqn:final-eeij-minus-bbij-2} \\
    {C}_{\ell}^{rot, {EB}, (ij)} - {C}_{\ell}^{rot, {EB}, (ji)}
      &\simeq 
        \sin(2\alpha_j - 2\alpha_i)
        e^{-4C^{\alpha}(0)} 
        \biggl[(\tilde{C}_{\ell}^{{E}{E}} + \tilde{C}_{\ell}^{{B}{B}})
        \nonumber\\
      &\quad +
        \sum_{\ell^\prime L} \frac{(2\ell^\prime + 1)(2L + 1)}{\pi} C_L^{\alpha\alpha}
        \begin{pmatrix}
            \ell & L & \ell^\prime \\
            2    & 0 & -2
        \end{pmatrix}^2
        (\tilde{C}_{\ell^{\prime}}^{{E}{E}} + \tilde{C}_{\ell^{\prime}}^{{B}{B}})\biggr],
      \label{eqn:final-ebij-minus-ebji-2} \\
    {C}_{\ell}^{rot, {EB}, (ij)} + {C}_{\ell}^{rot, {EB}, (ji)}
      &\simeq 
        \sin(4\beta + 2\alpha_i + 2\alpha_j)\ e^{-4C^{\alpha}(0)} 
        \biggl[(\tilde{C}_{\ell}^{{E}{E}} - \tilde{C}_{\ell}^{{B}{B}})
        \nonumber\\
      &\quad -
        \sum_{\ell^\prime L} \frac{(2\ell^\prime + 1)(2L + 1)}{\pi} C_L^{\alpha\alpha}
        (-1)^{\ell + L + \ell^\prime}
        \begin{pmatrix}
            \ell & L & \ell^\prime \\
            2    & 0 & -2
        \end{pmatrix}^2
        (\tilde{C}_{\ell^{\prime}}^{{E}{E}} - \tilde{C}_{\ell^{\prime}}^{{B}{B}})\biggr].
      \label{eqn:final-ebij-plus-ebji-2}
\end{align}

\end{document}